\documentclass[twocolumn,showpacs,preprintnumbers,amsmath,amssymb]{revtex4}

\usepackage{graphicx}
\usepackage{dcolumn}
\usepackage{bm}

\begin{document}

\title{Parallelization of Cellular Automata
for Surface Reactions}

\author{R. Salazar}
\email{R.Salazar@tue.nl}
\author{A.P.J. Jansen}
\email{tgtatj@chem.tue.nl}
\affiliation{
Schuit Institute of Catalysis (ST/SKA), Eindhoven University of
Technology, P.O. Box 513, 5600 MB Eindhoven, The Netherlands.}

\author{V.N. Kuzovkov}
\email{kuzovkov@latnet.lv}
\affiliation{
Institute for Solid State Physics, University of Latvia,
Kengaraga 8, LV-1063, Riga, Latvia.}

\date{\today}

\begin{abstract}
We present a parallel implementation of cellular automata
to simulate chemical reactions on surfaces. The scaling 
of the computer time with the number of processors for this parallel 
implementation is quite close to the ideal $T/P$, where $T$ is the computer
time used for one single processor and $P$ the number of processors. Two
examples are presented to test the algorithm, the simple A$+$B$\to0$ model 
and a realistic model for CO oxidation on Pt$(110)$.
By using large parallel simulations, it is possible to derive
scaling laws which allow us to extrapolate to even larger system sizes 
and faster diffusion coefficients allowing us to make direct comparisons 
with experiments.
\end{abstract}

\pacs{ 82.65.+r; 82.20.Wt; 02.70.Tt; 82.40.Np; 89.75.Da}

\keywords{Parallel Cellular Automata, Heterogeneous catalysis,   
Scaling laws}

\maketitle

\section{Introduction}

One of the most interesting features of surface reactions
is that in a large number of cases produce pattern formation, 
structures with some well--defined length scale, sometimes with symmetries 
and temporal behavior, such as oscillations, traveling waves,
spirals, Turing patterns, etc \cite{rab2000,imb1995}. 
A usual approach to study this
pattern formation is reaction--diffusion (RD) equations \cite{wal1997},
which simulate the dynamic behavior of chemical reactions on surfaces.
However, these partial
differential equations give only approximate solutions and in 
several cases completely wrong results, because they are based on 
the local mean field approximation, meaning well--mixed reactants 
at microscopic level, ignoring all the correlation terms between 
reactants locally, fluctuations and lateral interactions in the
adsorbate \cite{win2002}.
The RD equations describe the coverage which are macroscopic continuum 
variables which neglect the discrete structure of matter,
and do not describe the actual chemical process underlying pattern 
formation.
In fact from experimental studies \cite{win2002} it is known that
to model correctly, a modified kinetic has to be assumed different 
from the prescribed for RD.

Based in some general assumptions of the physics processes involved,
an esencial master equation can be derived that completely describes the 
microscopic dynamic of the system \cite{kam1981}. 
An exact method to solve this master equation is the Monte Carlo
(MC) method \cite{gil1977,jan1995,luk1998,zve2001}. In a Monte Carlo method a 
sequence of discrete events (reactions, including diffusion) is generated 
on a 2D lattice which represents the surface sites. These events 
are generated in general in a random way, looking for possible 
enabled reactions between nearest neighbors on the lattice, and 
doing the reactions with some probability in correspondence with 
some defined reaction rates. 

To compare MC simulations with experimental
pattern formation it is necessary to fill the gap between the length scale of
the individual particles and the diffusion length.
The regime where spatio--temporal patterns usually occurs, from 
$\mu$m to mm scales, is orders of magnitude
larger than the nm scale of individual particles. 
However, some new experiments \cite{win1997,vol1999} show that the 
fast kinetics processes 
are typically accompanied by the appearance of nanostructures.
Only microscopic simulations could deal with this two--scales behavior.
However this regime implies lattices sizes above $10^6 \times 10^6$
and very large values for the diffusion rates to produce agreement with
the experimental observed scales on pattern formation.
This would be a very large and slow simulation, due to the huge number
of particles involved and the fast diffusion rates which means that most 
of the simulation time is spend doing diffusion of particles instead of 
chemical reactions.
Fortunately, we do not need to simulate every time such a macroscopic system
or experimental diffusion rates. We only need to find scaling laws 
for lengths and diffusion coefficients, i.e from nm scale to $\mu$m
or mm scale, and from microscopic to real time. This is the possibility
which we explore in this paper by using large simulations within parallel
computers.

Although only MC simulations provides solutions to the exact master
equations for the surface reactions, they are not suitable for efficient 
parallelization due to the random selection of lattice sites used.
However, there is another important approach to simulate discrete 
events on lattices, the Cellular Automata (CA) 
\cite{tof1987,wei1997}. This approach is fully parallel 
in the sense that all the lattice sites can be updated simultaneously.
This has the advantages that fewer random numbers are required,
only those for the reaction probabilities (CA codes are faster than MC), 
and the global updating is easier to implement in a parallel code.

Under some well--defined conditions a CA is equivalent to a MC
simulation (see ref. \cite{kor1998} and section II). 
Then a CA could be used as
a MC simulation and its parallelization will provide with the
required scaling laws. 
In this paper we present for first time an attempt to provide a tool
to get these scaling laws.
The fact that CA are ideal simulation 
methods for parallelization is shown in
a recent special issue dedicated to CA in {\em Parallel Computing}
\cite{rev2000}.
In particular a quite interesting paper by J.R.~Weimar \cite{wei2000}
presents an object oriented parallel CA, also based in ref. 
\cite{kor1998}, and explores the possibility of divide the surface in 
regions to be simulated by RD or CA according to the level of detail 
required.

In this paper we will describe in detail the implementation of the
parallel version of a CA simulating
chemical reactions on surfaces, but the ideas discussed here are
also applicable to more general CA simulations.
In section II we will describe briefly the Cellular Automata method.
In section III we explain the key points to implement the parallelization
of the algorithm. In section IV we will present results of the application 
of the parallel algorithm to the A$+$B$\to0$ model, and to a realistic 
model of oxidation of CO on Pt$(110)$.
Finally in section V we will draw some conclusion.

\section{Cellular Automata}

A cellular automaton (CA) is a regular array of cells.
Each cell can be in one of a set of possible states.
The CA evolves in time in discrete steps by changing the 
states of all cells simultaneously.
The next state which a cell will take is based on the previous 
state of the cell and the states of some neighboring cells.
A CA is defined by providing prescriptions for
the lattice, the set of states, the neighbors, and the
transition rules \cite{tof1987,wei1997}.
The idea of CA deserves by itself a lot of study, since 
in general the evolution of a CA cannot be predicted other than 
by executing it. Additionally the amount of possible CAs is 
quite large. For instance, in the simplest one--dimensional case,
with two states and two neighbors, there are $256$ possible CA.

\begin{figure}
\includegraphics[width=6cm]{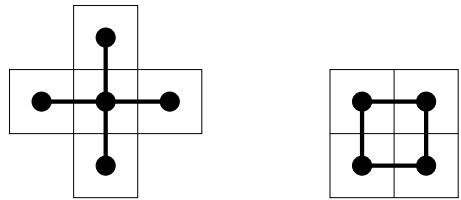}
\caption{
\label{margolus}
Von Neumann neighborhood (left) and Margolus neighborhood (right). 
The lines joining points show the nearest neighbor couples in each case.
}
\end{figure}

Our CA uses the standard square two--dimensional lattice 
of size $L\times L$.
The states of each cell represent occupation with some kind of particle. 
The Margolus neighborhood \cite{tof1987} is used
instead of the common von Neumann neighborhood. Both definitions of 
neighborhood are shown in Fig.\ref{margolus}. 
In order to obey the CA laws in von Neumann neighborhood, it is
necessary \cite{cho1988} disobey the laws of stoichiometry because
one particle could participate in more than one reactive pair. 
Similar problems arise with the diffusion of particles \cite{cho1991}.
The use of a Margolus neighborhood overcomes these difficulties 
\cite{mai1991,mai1992,mai1992b,mai1993}.
Using the value of the chemical rates we set up some 
probabilistic transition rules to change the states of the cells 
inside each Margolus block.

\begin{figure}
\includegraphics[width=8cm]{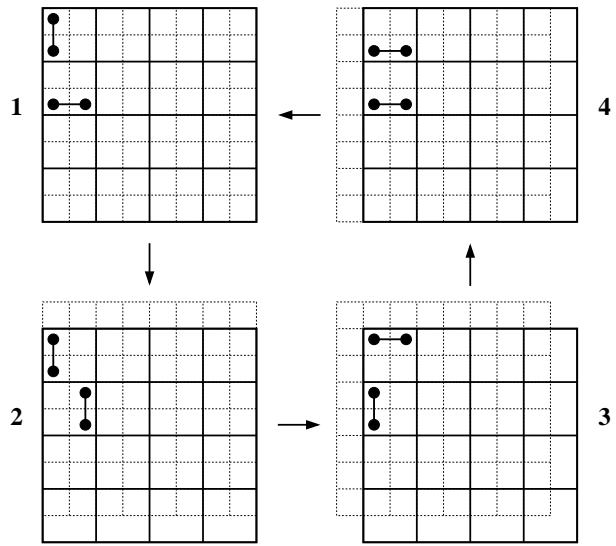}
\caption{
\label{tiling}
The four possible tilings using Margolus blocks. The arrows show the
sequence order.
The lines joining points illustrate the MC updating of pairs 
inside blocks.
}
\end{figure}

The Margolus blocks, Fig.\ref{margolus}, are used in the following way.
The blocks are periodically repeated to build up a tiled mask over the 
whole lattice, considering periodic boundary conditions. 
In this way there are four possible tilings as shown in 
Fig.\ref{tiling}.
Only neighbor sites belonging to the same
block can react, so all the blocks can be accessed in parallel.
Inside each block a Monte Carlo update is done. After a full lattice
update the whole procedure starts over again using another of the
four possible tilings. The dynamics is not confined to blocks, 
because the boundary between blocks is changing from one global sweep 
to the next.

For the four tilings shown in Fig.\ref{tiling}, we chose
randomly one of the four possible sequences of tilings:
($1$,$2$,$3$,$4$), ($2$,$3$,$4$,$1$), ($3$,$4$,$1$,$2$), and
($4$,$1$,$2$,$3$).
They show always the same clockwise cyclic sequence, but starting at a
different tiling. This produces better boundary diffusion and mixing between 
blocks.

A schematic description of the steps to implement this CA is given
at the following:
\begin{itemize}
\item[1. ] Choose randomly one of the four possible 
sequences of tilings: ($1$,$2$,$3$,$4$), ($2$,$3$,$4$,$1$), ($3$,$4$,$1$,$2$),
($4$,$1$,$2$,$3$).
\item[2. ] Choose consecutive tilings from the sequence of step 1.
\item[3. ] Sweep over all the blocks and in each block make a single
Monte Carlo update:
\begin{itemize}
\item[3.1. ] Choose randomly one pair of neighbors.
\item[3.2. ] Choose a reaction $i$ from the set of all the
possible reactions with a probability proportional to the reaction
rate $k_i$.
\item[3.3. ] Check if the reaction chosen in step 3.2 is possible on the
sites chosen in step 3.1. If it is possible do the reaction, 
otherwise do nothing.
\end{itemize}
\item[4. ] Increase the time by $\Delta t/4$, and return to step 2 until 
the sequence of tilings is completed.
\item[5. ] Return to step 1.
\end{itemize}

The updating scheme inside each block is the same as in the 
Dynamic Monte Carlo algorithm called Random Selection Method
\cite{luk1998,zve2001}.
Consequently, the time increment is the same as one Monte Carlo
Step (MCS) \cite{luk1998}:
\begin{equation}
\Delta t = \frac 1 {L^2 \sum_i k_i}
\end{equation}
where the sum in the denominator corresponds to the sum of all the
possible reaction rates.

There is a large number of possible CA prescriptions which could try to
reproduce a MC simulation of chemical reactions on surfaces. 
In fact, an extensive study made by J.~Mai 
\cite{mai1991,mai1992,mai1992b,mai1993}
shows that in general it is difficult produce good CAs reproducing
chemical reactions and diffusion.
However, in a recent paper Kortl\"uke \cite{kor1998} studies under
which conditions a CA could reproduce adequately a MC simulation
of chemical reactions on surfaces.
He found that the main requirement is using large diffusion coefficients.
It is required some sort of compromise between MC and CA:
it is necessary use a regular array of blocks as in CA, and a MC update
scheme has to be used inside each block.
The CA which we use here is a small modification of that used by Kortl\"uke
originally \cite{kor1998}. 
We use blocks of $4$--sites (Margolus blocks) instead of
$2$--sites (Hantels blocks).
The main advantage of using this $4$--sites instead of $2$--sites
is that reduces the level of CA noise (see \cite{kor1998}) inside 
each block.
This CA noise is the difference between the diffusion simulated with
CA and the correct diffusion simulated with MC.
In fact using larger blocks the noise will be smaller.

Additionally we present here a full realization of the CA parallel 
computing idea by implementing this in a parallel code as is shown in
the next section.
This is the main reason for use CA instead MC as a solution for large--time
and large--sizes simulations. Otherwise the only use of CA instead MC in 
a serial simulation does not speed up the simulation substantially, in 
the best cases only $\sim 10$--$20\%$. This means that the implementation of 
an efficient CA is very important.

Despite that we have to accept that our CA is an approximation to exact MC
microscopic simulations, we known when the approximation is good
and we can always check that the CA is reproducing the MC results. 
In this way we can think about the CA as the MC realization of the exact 
master equation for the chemical processes, getting in this way,
the possibility of obtain the scaling laws mentioned at the introduction
for these physical systems. In this paper we do not discus the problem
of quality of the CA approximation to the MC realization. This was done in 
\cite{kor1998} and we have made for our modified CA a similar study with 
similar results.

The fact that blocks can be updated without referring to other
blocks allows us to do CA as a parallel algorithm, because
the whole set of blocks can be updated at the same time.
This is the subject of the next section.

\section{Parallelization}

The aim of the parallelization of any code is to distribute the whole 
simulation over several computer processors, also called nodes. 
We define $speed(P)=1/T_P$, where $T_P$ is the computer time used by
$P$--processors to complete a simulation.
The optimum result is that this multiplies the speed of the simulation 
with the number of nodes, $speed(P)= P \times speed(1)$.
However, the usual result of parallelize a code is not the optimum.
The key point to achieve a good speed up is that the time spent by 
each node sending and receiving information from/to the other nodes is 
small in comparison with the time consumed doing computing in each node. 

\begin{figure}
\includegraphics[width=8cm]{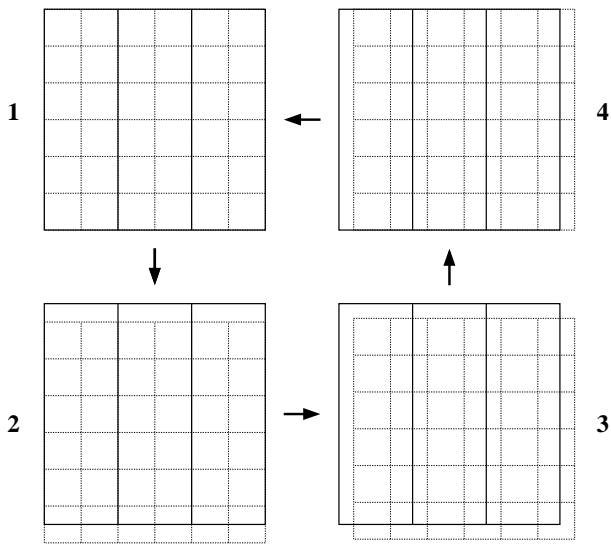}
\caption{
\label{strips}
Distribution of the lattice in strips sublattices for 
computing in each node. Division for each tiling
shown in Fig.\ref{tiling}. 
}
\end{figure}

In order to distribute the work, we use here a geometrical division 
as usual \cite{rev2000} in spatial extended systems, i.e., the full 
lattice is divide in sublattices of equal area. 
The time spent sending data is proportional to the length of 
the borders, $\sim L$, and the time doing computing is proportional 
to the system size $\sim L^2$.
We divide the lattice in strips as is shown in Fig.\ref{strips}.
Considering the global periodic boundary conditions, each node has
periodicity in one direction and in the other directions has to share 
information with only two neighbor nodes.
Another possibility is divide the lattice in squares. This choice is less
convenient. Each node has to share information with 4 nodes instead 2,
and it is only possible to use a perfect square number of nodes 
$P=4,9,16,\dots$. 
Using the strips sublattices requires sending a factor $4/\sqrt{P}$ 
less data than using
the square sublattices \cite{note2}, provided to have $P \le 16$.

Note that due to the periodicity in the vertical direction inside
each node, it is not necessary to interchange border information between nodes
when we pass from the tiling $1 \to 2$ and $3 \to 4$. It is only necessary
when we pass from $2 \to 3$ and $4 \to 1$. This reduces in half the
amount of communication data.
From Fig.\ref{strips} we see the direction in which the first column of 
data in each sublattice has to be
be sent and received in each node. In the change of tilings $2 \to 3$ the
data goes from left to right and in $4 \to 1$ from right to left.
Instead of shifting the data of each node horizontally we use the
first column for sending to or receiving from other nodes.
As a consequence, and to get interaction between sites in the
same block, we consider the first column neighbor of the last one.
For this Blocking CA, where the reactions are confined inside the blocks, 
this produces an effective periodicity in the horizontal direction.
This extra periodic boundary condition inside each sublattice, 
makes the final parallel code very similar to a full lattice 
implementation.

We use the SPMD (single program, multiple data) model to implement the 
parallel simulation, in which every node executes the same code using
different sublattice. 
Also, we use a main node, called node zero, dedicated additionally to 
distribute and collect the global information needed for the input/output 
of the simulation and other global computing required.
Every node executes the same code, and
when some special code has to be 
executed for only some nodes, is necessary use a node 
identifier number $p$. From the global periodic boundary conditions and
the shape of the sublattices, there is periodicity also in the sequence of
nodes: the node left with respect to node $p=0$ is node $p=P-1$ (P is 
the total number of nodes) and the node right with respect to node $p=P-1$ 
is node $p=0$.

In the following we present the CA code for a single node.
\begin{itemize}
\item[1. ] $p=0$: Chose randomly one of the four possible tiling
sequences: ($1$,$2$,$3$,$4$), ($2$,$3$,$4$,$1$), ($3$,$4$,$1$,$2$),
($4$,$1$,$2$,$3$). Send that choice to the other nodes.\\
$p>0$: Receive the information of which tiling sequences has been
chosen.
\item[2. ] $\forall p$: Choose consecutive tilings from the sequence 
of step 1.
\item[N1. ] $\forall p$: If the previous tiling was $1$ or $2$ and 
the new tiling is $3$ or $4$ then \\
send the first column to the left node $p-1$, \\
receive the data from right node $p+1$ and put it in the first column.
\item[N2. ] $\forall p$: If the previous tiling was $3$ or $4$ and 
the new tiling is $1$ or $2$ then \\
send the first column to the right node $p+1$ \\
receive the data from left node $p-1$ and put it in the first column.
\item[3. ] $\forall p$: Sweep over all the blocks and in each block make a 
single Monte Carlo update.
\item[4. ] $\forall p$: Increase the time by $\Delta t/4$, and return to 
step 2 until the tiling sequence is completed.
\item[5. ] $\forall p$: Return to step 1.
\end{itemize}

We have modified step $1$ and added two new steps $N1$ and $N2$, to 
interchange information between the nodes. The rest is basically the same as
the single processor code, but applied to the respective sublattices for each 
node. In order to avoid undesired correlations between different sublattices,
special attention should be given to a good random number generator, 
producing different sequence of random numbers within each node
\cite{knu1998}.

For implementation of the interprocess communication and synchronization,
the message passing interface MPI library has been selected, 
because it provides source portability to different kind of computers.
This library provide, amongst a set of specialized and complete communication 
routines, a so called six--basic set of routines for interprocessor 
communication:
initialization, termination, getting the set of nodes, getting its own node
number, sending data to other nodes, and receiving data from other nodes. 

There are computers with connection between nodes using giga--ethernet or 
several processors inside the same machine, sharing the same 
memory. 
These computers represent optimal environments to test parallel 
algorithms, i.e. high--end supercomputers like Cray T3E, and middle--end 
supercomputers like Silicon Origin 2000.
However, we will show in the next section that the results of running 
this parallel CA algorithm in a low--end Beowulf cluster of PCs connected 
only via fast--ethernet produce already almost the ideal speed up.

\section{Results}

The parallel algorithm was tested on our local cluster of PCs,
(17 Athlon 1.1Ghz/256Mb, fast--ethernet, Linux 2.4.18, MPICH 1.1.2).
From the results we see that the improvement of the performance 
using $P$ processors with respect to a single processor is almost the 
ideal, $speed(P) = P \times speed(1)$. 
In order to test the speed up of the algorithm, we will use two 
systems from surface catalysis: 
the A$+$B$\to0$ model and a model for CO oxidation on Pt$(110)$ surface 
\cite{kuz1998}, which has produced several important results 
\cite{kor1998b,kor1999,kuz1999,kor1999b}.

The A$+$B$\to0$ model has been studied for a long time 
\cite{ovc1978,kuz1988,kot1996,pri1997,mar1999,arg2001}
In the pioneering analitycal paper by Ovchinnikov and Zeldovich
\cite{ovc1978} it was shown for first time that the kinetic law of
mass action is violated in this model, producing incorrect results
when standard chemical kinetics is used. An illustrative case, also used
in our paper, is a situation with equal concentrations of 
both reactants $C_A=C_B=C$, where the standard kinetics predicts
an asymptotic behavior $C \propto t^{-1}$. This prediction correspond
to the mean--field approximation and it is only valid for high 
dimensional systems \cite{ovc1978}, ${\cal D}\ge$4.
In the low dimensional systems ${\cal D}<$4 with diffusion controlled 
processes it has been proved, using renormalization group
arguments \cite{lee1995}, that the correct asymptotic behavior is:
$C \propto t^{-{\cal D}/4}$. A qualitative agreement with this behavior 
was shown for first time using MC simulations in reference \cite{tou1983}.

The asymptotic law needs large simulation time
$t_{max}$. The diffusion length $\xi(t)=\sqrt{Dt}$ defines the pattern 
formation scale.  A simulation until $t=t_{max}$ needs a lattice 
length $L \gg \xi(t_{max})$, which correspond to a large 
simulation time of the order of $t_{max}L^2 \sim t_{max}^2$. 
This case provides a good example of large--time and 
large--size systems with pattern formation to test our
parallel algorithm described in the previous section.

In the corresponding lattice model we consider two kinds of 
particles A and B.
The only possible chemical reaction is desorption of AB, which happen
when two particles A and B are next to each other, creating two
empty sites. This process occurs with a rate constant $k$. 
Additionally the particles are allowed to diffuse with
rate $D$. This happen when a particle is next to an empty site.
We simulate the behavior produced for an initial condition without
empty sites and where the same number of A and B particles are 
initially randomly distributed on the surface: $N_A=N_B=L^2/2$.

\begin{figure}
\includegraphics[width=8cm]{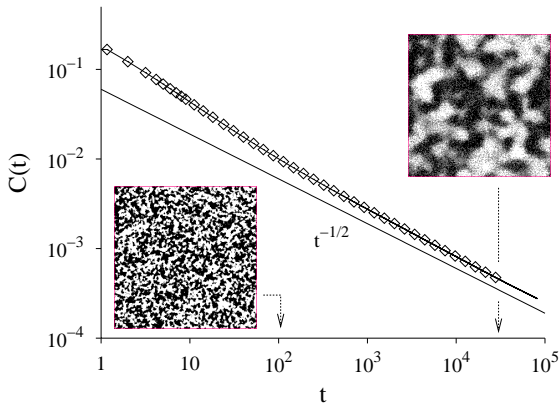}
\caption{
\label{fig1}
Temporal behavior of the concentration of particles for the 
A$+$B$\to0$ model starting with randomly mixed A and B.
The solid line is the result from the simulation using $16$
processors and the dots are from using a single processor.
The $t^{-1/2}$ curve shows the asymptotic behavior.
The two snapshots show the system at different times.
}
\end{figure}

\begin{figure}
\includegraphics[width=8cm]{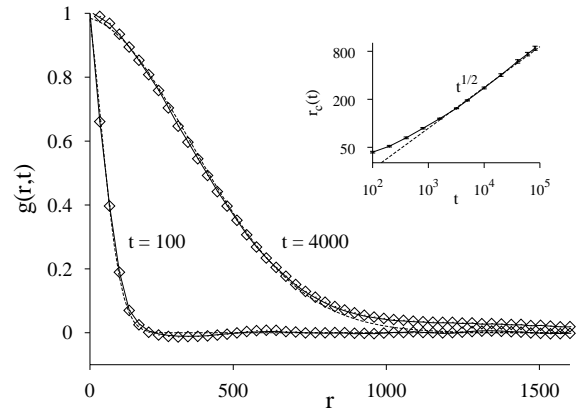}
\caption{
\label{fig2}
The radial correlation function for the A$+$B$\to 0$ model
The same times as the snapshots in Fig.\ref{fig1}.
The lines are the $16$ processor results and the dots the 
single processor results.
The dashed lines are fits with $\exp[-(r/r_c)^2]$.
The respective values $r_c(t)$ are plotted in the insert.
}
\end{figure}

In Fig.\ref{fig1} we present the temporal behavior and also illustrate 
the segregation process forming regions with high 
concentration of A or B, which increase in size with time. 
Also we show how the global concentration, $C=N_A/L^2=N_B/L^2$, 
diminishes with time following the asymptotic power--law 
$C \sim t^{-\frac 1 2}$. 
The system size used is $L=8192$ and the parameters are $k=D=1$. 
We present two sets of data in Fig.\ref{fig1}. 
The points correspond to a single
processor simulation, and the solid line corresponds to a parallel simulation 
using $16$ processors. Both simulations start with identical random
initial distribution of particles. It is noticeable that this initial 
distribution almost completely determines the following behavior of the system.
The snapshots shown as insert in Fig.\ref{fig1} are from the
parallel simulation. They are very similar to the ones obtained from the single
processor simulation, which uses the same initial conditions but 
a different sequence of random numbers to simulate the dynamics.
Moreover, in order to check quantitatively the agreement between 
spatial structures
between the single and parallel simulations, we present in Fig.\ref{fig2} 
the radial correlation function. Again the points correspond 
to the single processor case and the solid line to the parallel case. 
A correlation length $r_c$ could be obtained by fitting these correlation 
functions with $\exp[-(r/r_c)^2]$, which we show in Fig.\ref{fig2} 
by dashed lines. The obtained values of $r_c$ are plotted in the
insert. This shows that the correlation length for this dynamic also
follows a power--law $r_c \sim t^{\alpha}$. By numerical fitting we 
obtain $\alpha = 0.5122 \pm 0.012$.
An analytical asymptotical solution for this correlation functions
is given in \cite{kot1996}, $\exp(-r^2/4Dt)$ or $\exp[-c(r/\xi(t))^2]$.
The previous value obtained for $\alpha$ means that the diffusion length
$\xi(t)=\sqrt{Dt}$ define the scale of pattern formation for this
reaction model.

\begin{figure}
\includegraphics[width=8cm]{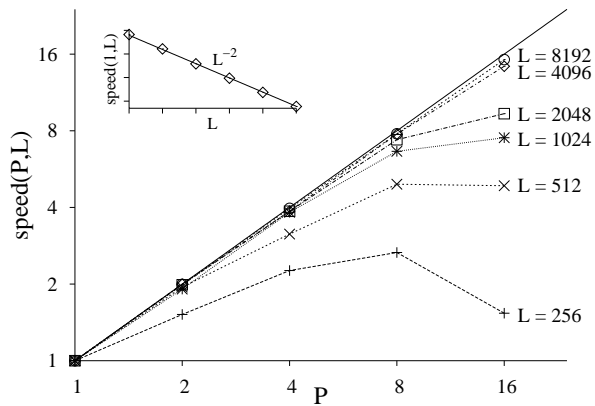}
\caption{
\label{fig3}
The speed of the simulation using $P$ processors normalized to the 
speed of one single processor for the A$+$B$\to 0$ model.
 Different system sizes 
$L=256,512,1024,4096,8192$ and number of processors $P=1,2,4,8,16$.
In the insert the behavior of the speed of one single processor.
}
\end{figure}

The performance or speed up of the parallel algorithm
for the A$+$B$\to 0$ model is shown in Fig.\ref{fig3},
using different system sizes $L=256,512,1024,4096,8192$ 
and number of processors $P=1,2,4,8,16$.
The simulated time for each computation was set to $t_{max}=500$.
The speed was normalized to the speed of the single processor case.
In the insert we can see the behavior of the single processor speed 
for each system sizes.
The advantage of use a large number of processors increases when the
system size increases, as we expect from the discussion in the previous 
section. 

The second model used to test the parallel algorithm is a model 
for CO oxidation on Pt$(100)$ and Pt$(110)$ surfaces 
\cite{kuz1998,kor1998b,kor1999,kuz1999,kor1999b}.  
This system shows different types of kinetic oscillations. 
On Pt$(100)$ local, irregular oscillations occur in a wide parameter 
interval, whereas on Pt$(110)$ globally synchronized oscillations exist 
only in a very narrow parameter interval. Both surfaces exhibit an 
$\alpha \rightleftharpoons \beta$ surface reconstruction, where $\alpha$
denotes the {\it hex} or $1 \times 2$ phase on Pt$(100)$ or Pt$(110)$,
respectively. $\beta$ denotes the unreconstructed $1 \times 1$ phase in 
both cases. Both surfaces have qualitatively quite similar properties 
with the exception of the dissociative adsorption of O$_2$. The ratio
of the sticking coefficients of O$_2$ on the two phases is 
$s_{\alpha}:s_{\beta} \approx 0.5:1$ for Pt$(110)$ and 
$s_{\alpha}:s_{\beta} \approx 10^{-2}:1$ for Pt$(100)$ \cite{imb1995}. 
From the experiments \cite{imb1995} it is known that kinetic
oscillations are closely connected with the 
$\alpha \rightleftharpoons \beta$ reconstruction of the Pt surfaces.

In the model \cite{kuz1998,kor1998b,kor1999,kuz1999,kor1999b}, 
CO is able to absorb onto a free surface
site with rate constant $y$ and to desorb from the surface with rate 
constant $k$, 
independent of the surface phase to which the site belongs.
O$_2$ adsorbs dissociatively onto two nearest neighbor sites with
rate constant $(1-y)s_{\chi}$ with $\chi = \alpha$,$\beta$. 
In addition CO is able to diffuse via hopping onto a 
vacant nearest neighbor site with rate constant $D$. The CO$+$O reaction 
occurs with rate constant $R$, when CO and O are in nearest neighbor sites
desorbing the reaction product CO$_2$.
The $\alpha \rightleftharpoons \beta$ surface phase transition is modeled as a
linear front propagation induced by the presence of CO in the border 
between phases with rate constant $V$.
Consider two nearest neighbor surface sites in the state $\alpha \beta$.
The transition $\alpha \beta \to \alpha \alpha$ 
($\alpha \beta \to \beta \beta$) occurs if none (at least one) of 
these two sites is occupied by CO.
Summarizing the above transition definitions written in the more usual
form of reaction equations gives:
\begin{eqnarray*}
&& \mbox{CO(g)} + S^{\chi} \rightleftharpoons \mbox{CO(a)}, \\ 
&& \mbox{O$_2$(g)} + 2S^{\alpha} \to 2\mbox{O(a)}, \\
&& \mbox{O$_2$(g)} + 2S^{\beta} \to 2\mbox{O(a)}, \\
&& \mbox{CO(a)} + S^{\chi} \to S^{\chi} + \mbox{CO(a)}, \\
&& \mbox{CO(a)} + \mbox{O(a)} \to \mbox{CO$_2$(g)} + 2S^{\chi}, \\
&& S^{\alpha} \rightleftharpoons S^{\beta},
\end{eqnarray*}
where $S$ stands for a free adsorption site, $\chi$ stands for either
$\alpha$ or $\beta$ and (a) or (g) for a particle adsorbed on the 
surface or in the gas phase, respectively. For additional
details see ref. \cite{kuz1998}.

Amongst several successful results of this model, we can mention
that it was one of the first microscopic models for CO oxidation on Pt
including surface 
reconstruction, which is nowadays widely accepted as the key element
in order to get oscillatory behavior. This model reproduce correctly
oscillatory regimes for both surfaces Pt$(100)$ and Pt$(110)$, by
changing only one parameter $s_{\alpha}$.
The diffusion of CO is consider explicitly and could be applied to
the fast diffusion regime without modification.
With this model an alternative mechanism for
global synchronization of oscillation has been suggested
\cite{kor1999}, different from the traditional
gas--phase coupling. This new mechanism is stochastic resonance,
obtained by including a spontaneous nucleation of one surface phase
in the other $\alpha \rightleftharpoons \beta$ at very low rates. 
One unique result reproducing experimental observations is
the transition into chaotic behavior via the Feigenbaum route
or period doubling \cite{kor1998b}.
It is also for this model that the compatibility of both microscopic
simulations MC and CA has been study extensively \cite{kor1998}. 

\begin{figure}
\includegraphics[width=8cm]{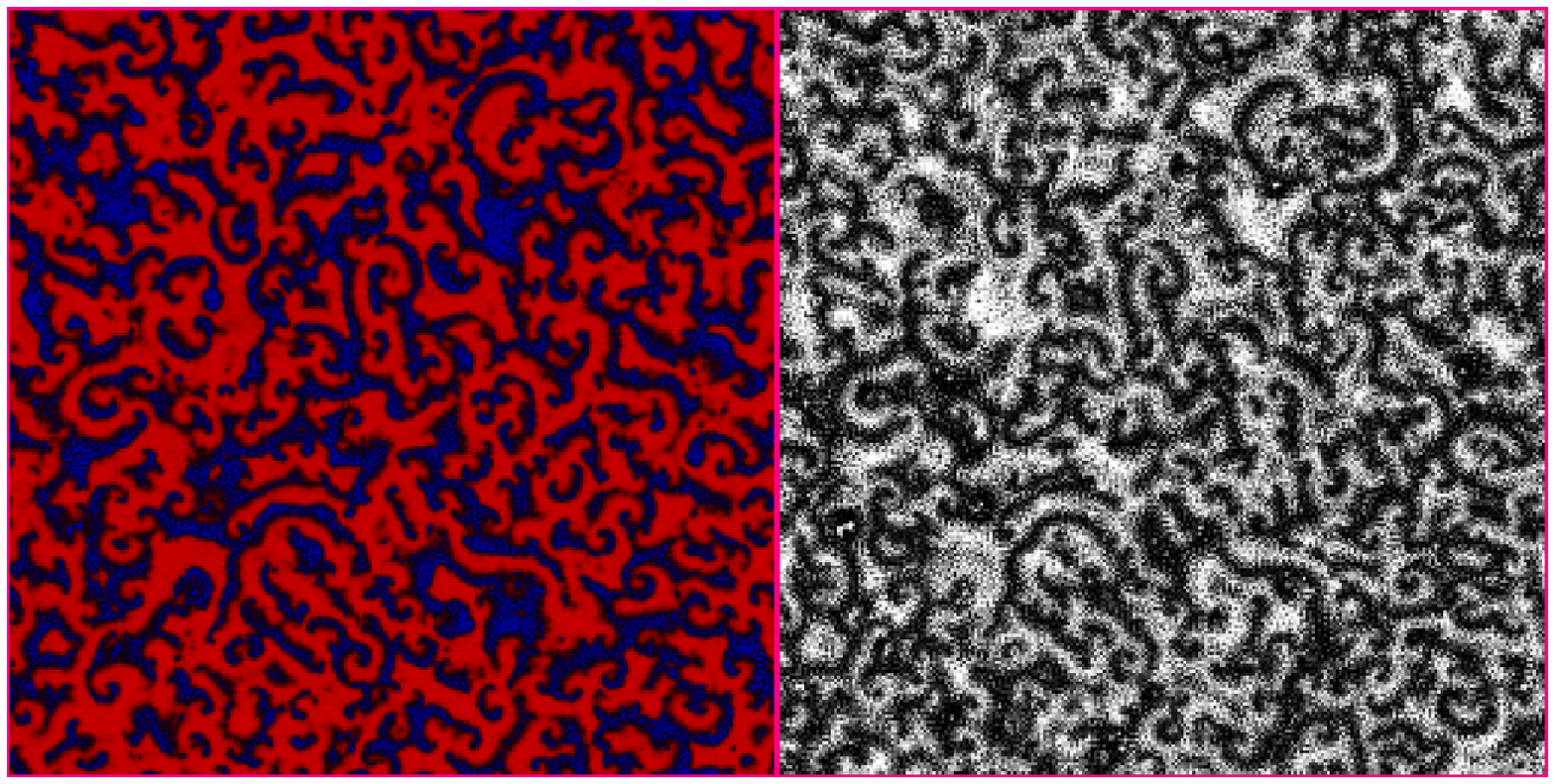}
\includegraphics[width=8cm]{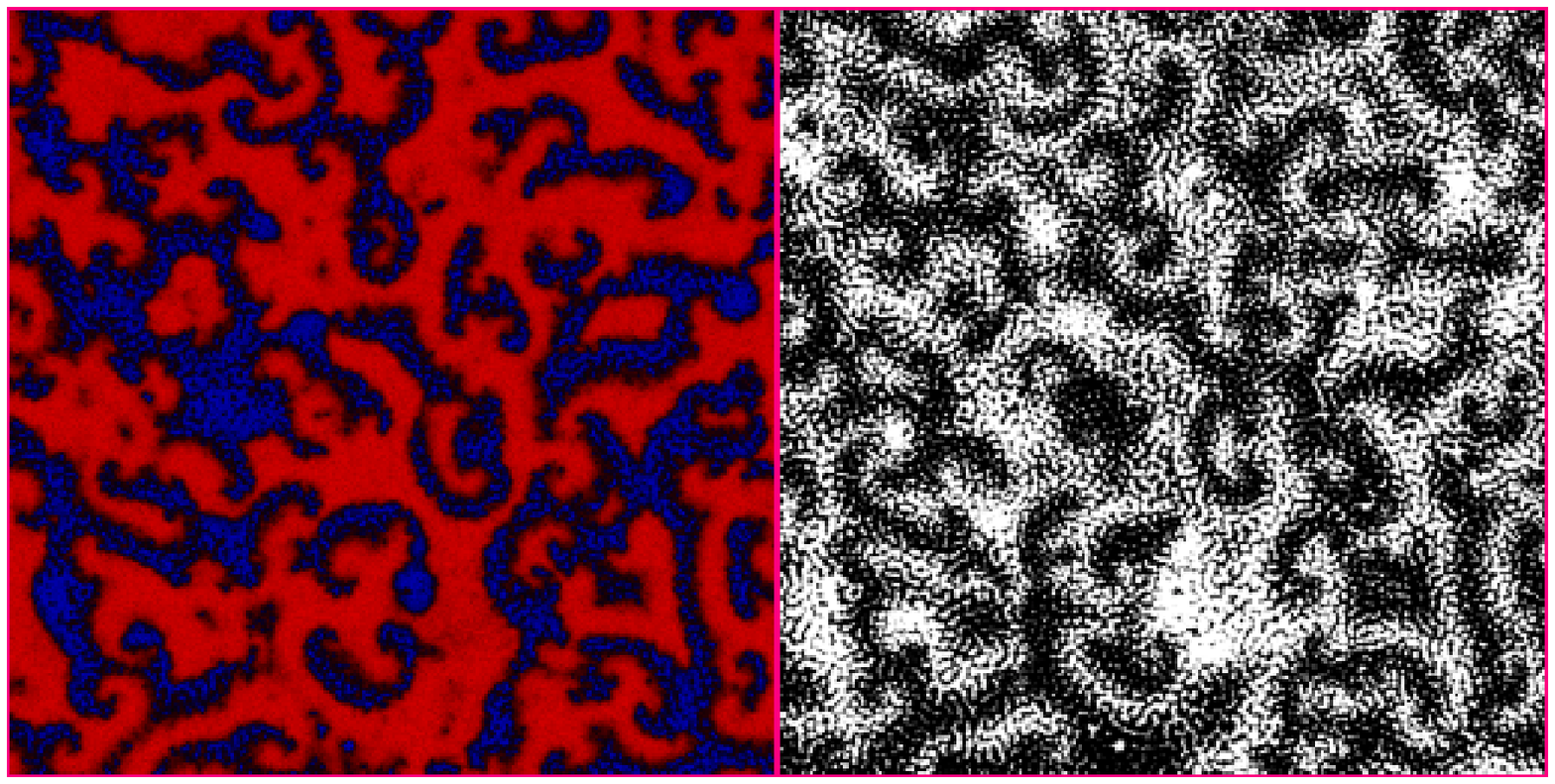}
\includegraphics[width=8cm]{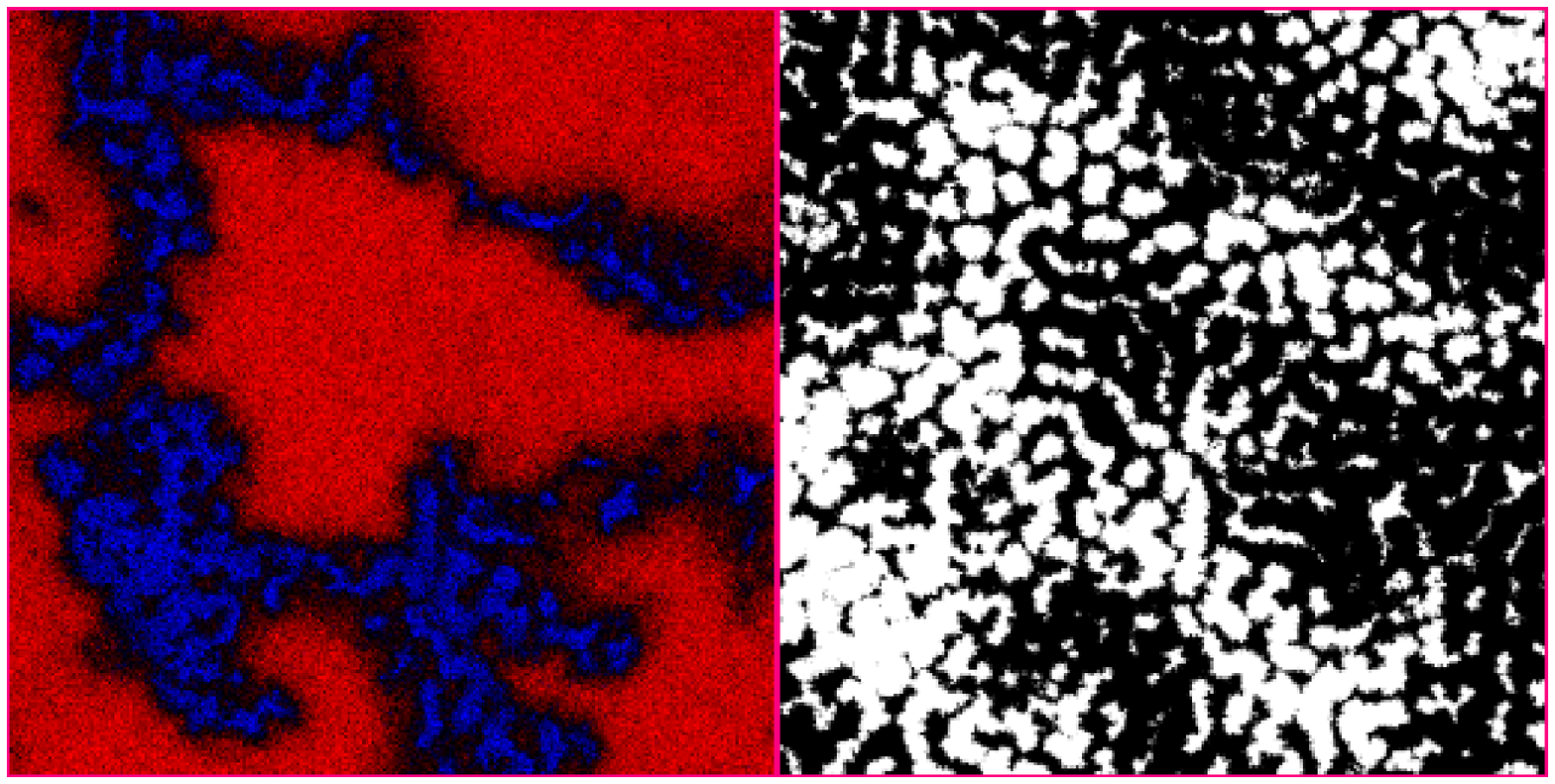}
\includegraphics[width=8cm]{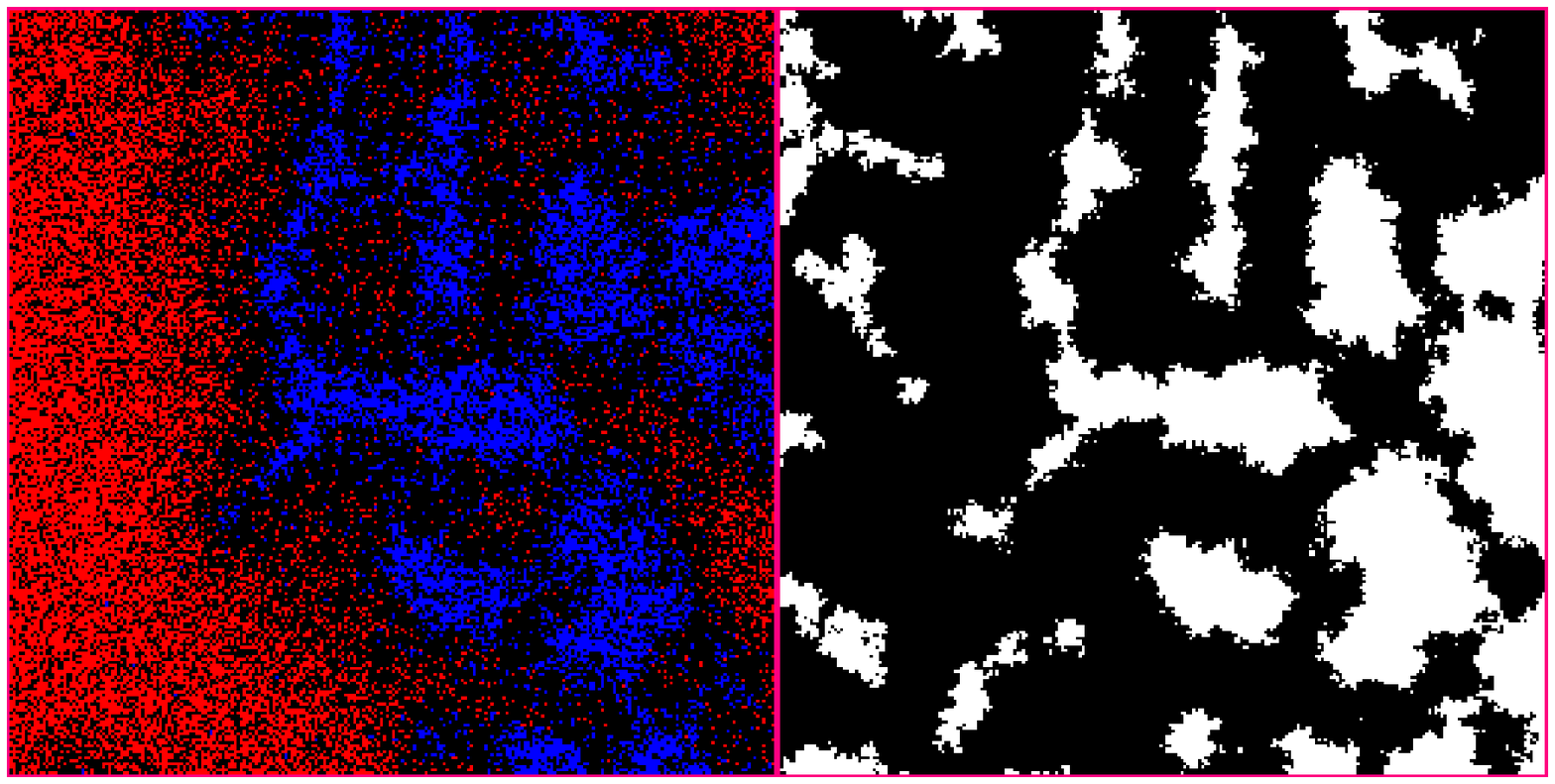}
\caption{
\label{fig5}
Sequence of snapshots of the model of CO oxidation on Pt$(110)$.
The left part shows the chemical species: CO particles are dark--grey,
O particles are light--grey, and empty sites are black. The right part
shows the structure of the surface: $\alpha$ phase sites are black, and
$\beta$ phase sites are white.
The parameters are $L=8192$, $D=250$, and $V=1$, $y=0.494$, $k=0.1$, $R=D$.  
From top to bottom, we show sections from the upper--left corner 
with sizes $4096 \times 4096$, $1024 \times 1024$, $256 \times 256$.
}
\end{figure}

In Fig.\ref{fig5} we show snapshots from a simulation
on Pt$(110)$ and a system size $L=8192$ using the parameter values,
$D=250$, $V=1$, $y=0.494$, $k=0.1$, $R=D$. In the left part
we plot the chemical species: CO particles are dark--grey,
O particles are light--grey, and empty sites are black.
The right part shows the structure of the surface: $\alpha$ phase 
sites are black, and $\beta$ phase sites are white.
The pattern formation in this regime shows a spatio--temporal behavior, 
where a spiral dynamics is the dominant phenomena. 
It is interesting see the different
structures at different spatial scales. For this purpose
we include in Fig.\ref{fig5} sections from the upper--left corner
with sizes $4096 \times 4096$, $1024 \times 1024$, $256 \times 256$,
from top to bottom.
This sequence shows that the spiral dynamics occurs on a
slowly varying island structure of sizes of the order of $\sqrt{D/V}$.
The fact that we can see both mesoscopic and microscopic pattern formation 
is a quite interesting feature of the model, which has some experimental
evidence \cite{win1997,vol1999} and has been studied theoretically in
\cite{hil2002} by including lateral interactions between adsorbed particles.
The model used here is simpler, because does not need that consideration
in order to obtain nanostructures.

\begin{figure}
\includegraphics[width=8cm]{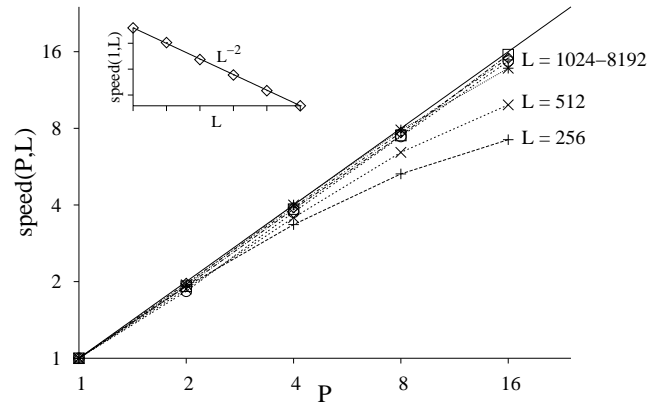}
\caption{
\label{fig4}
The same as Fig.\ref{fig3}, but for the model of CO
oxidation on Pt$(110)$.
}
\end{figure}

In Fig.\ref{fig4} we analyze the speed up of the parallel 
algorithm of this realistic model. We use system sizes,
$L=256,512,1024,4096,8192$ and number of nodes $P=1,2,4,8,16$.
The simulated time for each computation was also set to $t_{max}=500$.
Here we can see that the speed up of the parallel algorithm is
good even for small system sizes.
This is because the amount of computing
in each node for this model is larger than for the A$+$B$\to0$ model, 
while the amount of communication data is the same in both models.

\section{Conclusions}

In this paper we present a tool to obtain scaling laws
connecting experimental system sizes and diffusion coefficients to
standard values in microscopic MC simulations. By using a CA
equivalent to MC simulation we provide an efficient parallelization
algorithm. We have explained in detail how to implement the parallelization.
The speed up of the algorithm is almost ideal and it is much better
for larger system sizes and more complex models.
A full description and analysis of the scaling laws for the second
model used here is in preparation \cite{sal2002}.

\begin{acknowledgments}
We thank J.J.~Lukkien and S.~Nedea for stimulating discussions.
This work was supported by the Nederlanse Organisatie voor 
Wetenschapperl\"yk Onderzoek (NWO), and the EC Excellence Center of 
Advanced Material Research and Technology (contract N  1CA1-CT-2080-7007).
We thank the National Research School Combination Catalysis (NRSCC)
for computational facilities.
\end{acknowledgments}

\newpage


\begin{thebibliography}{3}
\bibitem{note2} By using $P$ processors and a system size $L^2$
the total interface between blocks is for the squares
$2 L \sqrt{P}$ and for the strips $L P$. However, due
to the cyclic tiling sequence shown in Fig.\ref{strips} we
reduce the amount of data to be send in half for the strips case.
The ratio of data to be sent between squares and strips then is:
$(2L\sqrt{P})/(LP/2)=4/\sqrt{P}$.
\end{thebibliography}
\end{document}